\documentclass[]{spie}  

\usepackage{amsmath,amsfonts,amssymb}
\usepackage{graphicx}
\usepackage[colorlinks=true, allcolors=blue]{hyperref}

\usepackage{mathtools}

\title{Seidr update: photonic ‘black magic’ for high-contrast interferometry using kernel-nulling and photonic lanterns}

\author[a]{Nathan~K.~Long}
\author[a]{Daniel~S.~Dahl}
\author[b]{Christopher~H.~Betters}
\author[a]{Julia~J.~Bryant}
\author[c]{Nick~Cvetojevic}
\author[d]{Michael~J.~Ireland}
\author[e]{Stefan~Kraus}
\author[b]{Sergio~Leon-Saval}
\author[c]{Frantz~Martinache}
\author[c]{Marc-Antoine~Martinod}
\author[a,b]{Barnaby~Norris}
\author[e]{Jyotirmay~Paul}
\author[a]{Akira~Rodziewicz-Ryan}
\author[a,f]{Adam~K.~Taras}
\author[b]{Jin~Wei}
\author[a,b]{Peter~G.~Tuthill}
\affil[a]{Astralis-USyd, Sydney Institute for Astronomy (SIfA), School of Physics, University of Sydney, Camperdown, NSW, Australia}
\affil[b]{Sydney Astrophotonics Instrumentation Laboratories, Sydney Institute for Astronomy (SIfA), University of Sydney, Camperdown, NSW, Australia}
\affil[c]{Observatoire de la Cote d’Azur, Université Côte d’Azur, Nice, France}
\affil[d]{Astralis-AITC, Research School of Astronomy and Astrophysics, Australian National University, Canberra, ACT, Australia}
\affil[e]{Astrophysics Group, University of Exeter, Exeter, United Kingdom}
\affil[f]{Leiden University, Leiden, Netherlands}

\authorinfo{N.K.L. e-mail: nathan.long@sydney.edu.au}

\begin{document} 
\maketitle

\begin{abstract}
Seidr is a new interferometric beam combiner within the Asgard Suite, utilizing infrastructure common to the BIFROST instrument at the Very Large Telescope Interferometer. Seidr combines hybrid mode-selective photonic lantern injection modules with a kernel-nulling photonic chip backend to enable deep H-band nulling for high-contrast studies of exoplanets, exomoons, and circumstellar dust. This instrument update summarizes Seidr\textquoteright s current design maturity and recent simulations of the point source – to – lantern outputs. We also outline progress on our neural network–based wavefront estimation scheme, which uses the photonic lantern outputs to sense phase fluctuations, designed to feed back to Baldr\textquoteright s deformable mirror, and improve nuller light injection.

\end{abstract}


\section{INTRODUCTION}
\label{sec:intro}  

Seidr is a kernel-nulling instrument within the Asgard Instrument Suite~\cite{Taras2024_seidr, Martinod2023, Taras2024_heimdallr, Courtney2024_baldr, Garreau2024_nott}, which operates at the Very Large Telescope Interferometer (VLTI), providing a world-class testbed for kernel-nulling technology. The VLTI supports both long baselines (up to 200~m), enabling high angular resolutions, and large telescope apertures (up to 8~m), enabling faint target detection~\cite{Abuter2026}. ``Seidr'' is a term in old Norse, which translates to a form of black magic --- a fitting term for the operation of a photonic nulling device.

The development of high contrast and high angular resolution optical instrumentation allows the innermost regions of extrasolar systems to be imaged, where nulling interferometry is able to destructively combine starlight from multiple apertures, outputting the remaining off-axis light (e.g.~\cite{Martinod2021}). Kernel-nulling introduces an additional stage to classical nulling, constructing kernel observables which are robust to second-order phase fluctuations~\cite{Martinache2018_kernel}. 

There are several potential science objectives of Seidr. Firstly, the direct imaging of young, hot Jupiter-sized exoplanets, which are self-luminous. This in particular complements NOTT, an Asgard suite instrument\textquoteright s longer wavelength capabilities, to fill a larger proportion of the hot Jupiter parameter space~\cite{Taras2024_seidr}. Secondly, the direct imaging of dusty circumstellar environments, composed of both scattered and thermal light, which could potentially reveal the existence of hidden off-axis structures. Thirdly, we envision a mature iteration of Seidr could potentially directly image warm exomoons around young, large planets, which would represent a giant leap forward in extrasolar system understanding.

Seidr uses hybrid mode-selective lanterns (HMS-PLs) for optimal light injection into a kernel nulling chip. Photonic lanterns are fibre based photonic devices, which transform a point-spread function (PSF) electric field, at a multi-mode fiber (MMF) input, to multiple single-mode fiber (SMF) output cores, via an adiabatic tapered transition, allowing spatial mode sorting of the ``lantern modes''~\cite{Leon2013}. HMS-PLs are a special type of PL that are specifically designed to guide the fundamental mode into a single mode-selective core, while the higher order modes are guided to the rest of the wavefront sensing cores~\cite{Norris2022_hybrid}. HMS-PLs overcome a major limitation of alternative wavefront sensing technologies; wavefront sensing light and science light exist on the same path, reducing wavefront errors associated with non-common path aberrations.

Four HMS-PLs are introduced in Seidr, one for each beam of the VLTI. The light in the mode-selective cores of each lantern is then injected into the kernel-nulling chip, while the wavefront sensing core intensities are input into a wavefront correction algorithm, designed to estimate a wavefront correction to apply to an upstream deformable mirror. The objective of the wavefront correction algorithm is to maximize the amount of light injected into the mode-selective core, thus maximizing the light injected into the nulling chip.

Machine learning (ML) is well suited to the problem of mapping the non-linear relationship between the lantern core intensities and the pupil-plane wavefront estimations. For example, \cite{Norris2020}~shows neural network (NN)-based wavefront estimations for a 19-core photonic lantern. We develop a transformer NN as the wavefront estimation algorithm for HMS-PLs, focusing on its effect on the ratio of power injected into the kernel-nulling chip.

In this work, we provide an instrument update for Seidr, where we focus on how Seidr complements and interacts with the Asgard Suite in Section~\ref{fig:asgard}, the theory and simulation of HMS-PLs in Section~\ref{sec:pl}, our wavefront estimation algorithm in Section~\ref{sec:wfc}, the theory of kernel-nulling in Section~\ref{sec:knc}, and concluding remarks in Section~\ref{sec:concl}.

\section{SEIDR AND THE ASGARD SUITE} \label{sec:asgard}

Seidr is one of several instruments in the Asgard Instrument Suite~\cite{Martinod2023, Taras2024_heimdallr, Taras2024_seidr, Courtney2024_baldr, Garreau2024_nott}, designed for the European Southern Observatory\textquoteright s VLTI, at Mt. Paranal, Chile. 

The Asgard Instrument Suite is a next-generation instrument suite, aiming to advance high angular resolution imaging via advances in sensitivity, spectral resolution, and nulling interferometry. The VLTI has four unit telescopes (U1 to U4), which each have an 8~m diameter, and are positioned at the coordinates given in Table~\ref{tab:vlti_coords}, as well as four 1.8~m auxiliary telescopes. 

\begin{table}[ht]
    \caption{VLTI coordinates.}
    \label{tab:vlti_coords}
    \begin{center}
        \begin{tabular}{|c|c|c|}
            \hline
            \rule[-1ex]{0pt}{3.5ex} \textbf{Station} & \textbf{East} & \textbf{North} \\ \hline
            \rule[-1ex]{0pt}{3.5ex} U1 & -9.925 & -20.335 \\ \hline
            \rule[-1ex]{0pt}{3.5ex} U2 & 14.887 & 30.502 \\ \hline
            \rule[-1ex]{0pt}{3.5ex} U3 & 44.915 & 66.183 \\ \hline
            \rule[-1ex]{0pt}{3.5ex} U4 & 103.306 & 44.999 \\ \hline
        \end{tabular}
    \end{center}
\end{table}

Seidr is designed for H-band kernel-nulling at the VLTI to achieve very small angular separation and high contrast imaging, which is difficult for conventional interferometric visibility measurements or longer-wavelength nullers~\cite{Taras2024_seidr}. The light injected into the kernel-nulling chip is stabilized using the wavefront sensing capabilities of four HMS-PLs, one for each of the four telescopes, which feed back to an upstream deformable mirror (DM).

In addition to Seidr, the Asgard Instrument Suite is comprised of the following five instruments:

\begin{itemize}
    \item \textbf{Heimdallr:} Beam correction, fringe tracking, common detection. Passes information onto other instruments (generally not for science, though can be used for science).
    \item \textbf{Baldr:} Zernike wavefront sensor (not for science).
    \item \textbf{BIFROST:} High resolution spectrograph (for science).
    \item \textbf{NOTT:} Nuller (for science).
    \item \textbf{Solarstein:} Laser star for calibration and alignment (not for science).
\end{itemize}

\begin{figure}[ht] 
    \begin{center}
    \includegraphics[scale=0.35]{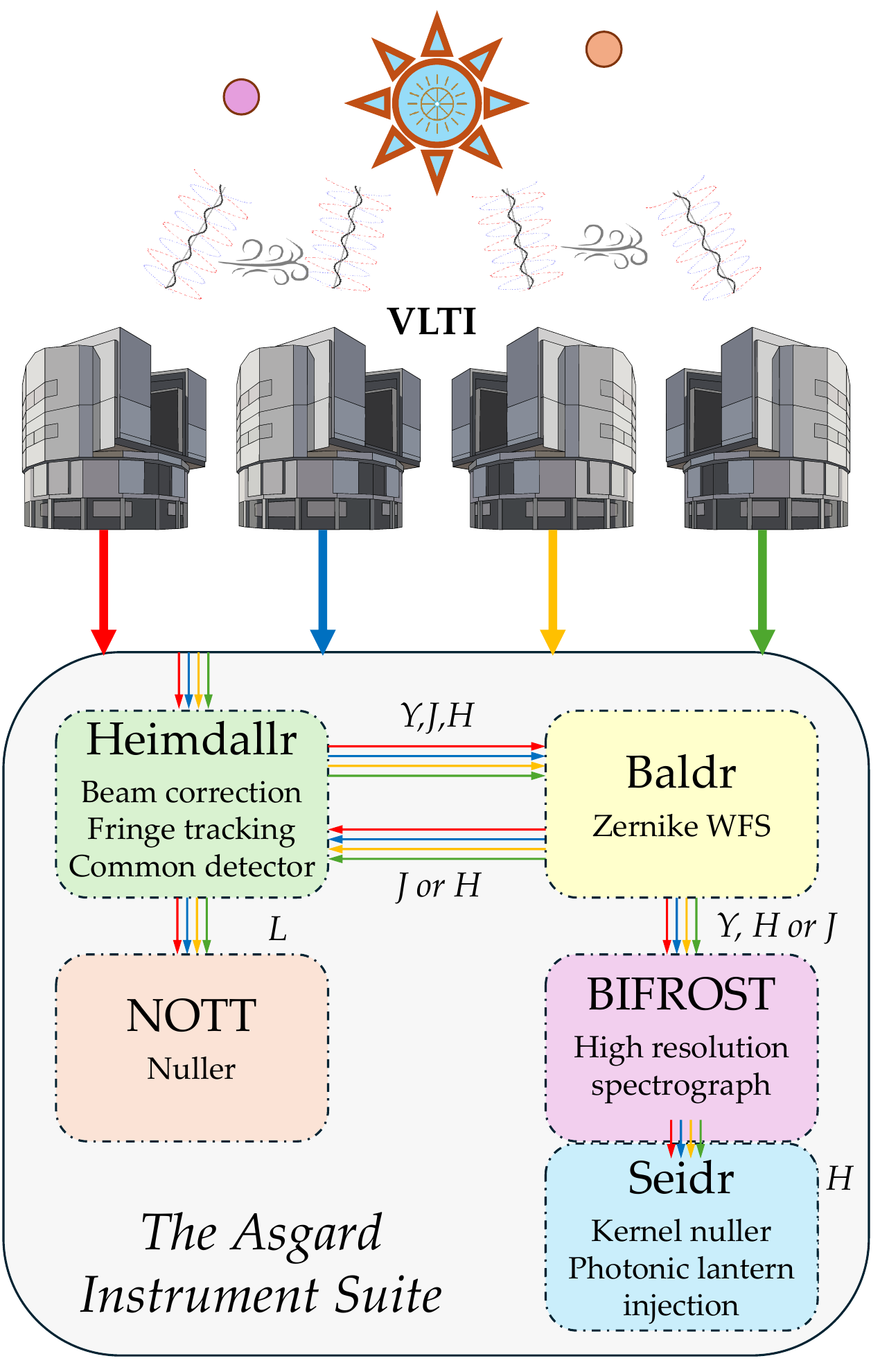}
    \end{center}
    \caption{The Asgard Instrument Suite within the Very Large Telescope Interferometer (VLTI).}
    \label{fig:asgard}
\end{figure}

A schematic the Asgard Instrument Suite is given in Figure~\ref{fig:asgard}, where the basic layout of the individual instruments and the direction of light propagation is shown (excluding Solarstein). 

Heimdallr is the gateway instrument to Asgard, providing fringe-tracking, low-order wavefront correction, and passing light to the rest of the instruments~\cite{Taras2024_heimdallr}. Baldr uses a Zernike wavefront sensor and a BMC Multi-3.5 DM to apply higher-order wavefront corrections to each of the four beams, which are then passed back to Heimdallr, as well as onto BIFROST~\cite{Courtney2024_baldr}. 

Seidr itself sits within BIFROST, hijacking its injection optics and C-RED One camera during operation. The H-band light from the four beams is fed into the four HMS-PLs, where only the light from the central mode-selective cores is injected into the kernel-nulling chip, while the light from the five wavefront sensing cores bypasses the chip straight to BIFROST\textquoteright s camera. The intensity measurements from the wavefront sensing cores are passed into the wavefront estimation algorithm (described in Section~\ref{sec:wfc}), which outputs an estimate of the wavefront correction used to actuate Baldr\textquoteright s DMs. Note that the bright output from the kernel-nulling chip is also passed into the wavefront estimation algorithm (simply included as one of the generic lantern outputs in this work).

The stability of Seidr\textquoteright s kernel nulls is, thus, dependent on Heimdallr for fringe tracking, Baldr for upstream wavefront correction and integration of Seidr\textquoteright s wavefront estimation capabilities, and BIFROST infrastructure and injection optics. Within the Asgard Instrument Suite, Seidr complements NOTT by shifting nulling science into the near-infrared, where the VLTI\textquoteright s angular resolution is higher and the HMS-PL injection helps improve stability and calibration.

\section{HYBRID MODE-SELECTIVE PHOTONIC LANTERN}
\label{sec:pl}

HMS-PLs are designed to optimize the injection of the fundamental LP01 mode into a single core for subsequent injection into a science instrument, which we define as the \textit{mode-selective core}~\cite{Norris2022_hybrid}. The light from the higher-order guided modes propagates into the rest of the cores, which is then used to estimate wavefront corrections to apply to an AO system, which we define as the \textit{wavefront-sensing cores}. In the case of Seidr, the light from the mode-selective core is injected into the kernel-nulling chip, and the intensities of the light from the wavefront sensing cores are input into an NN, which estimates a wavefront correction to apply to the upstream DM on Baldr.

\begin{figure}[ht] 
    \begin{center}
    \includegraphics[scale=0.9]{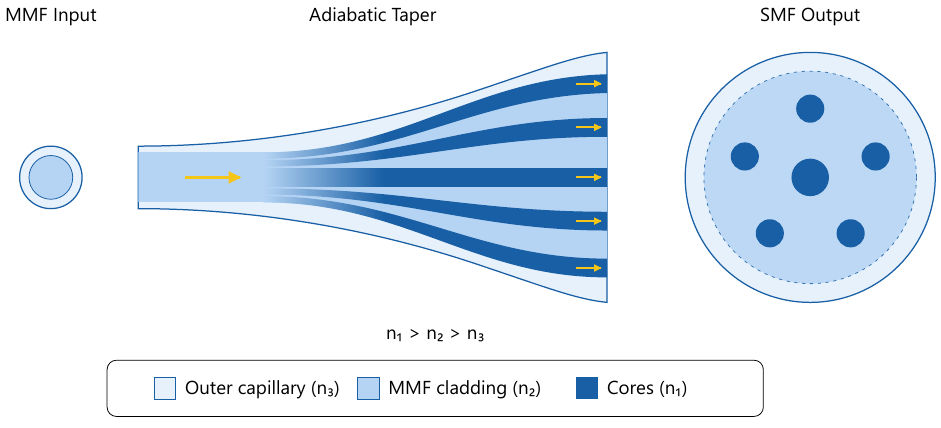}
    \end{center}
    \caption{Schematic of a 6-core hybrid mode-selective photonic lantern.}
    \label{fig:pl}
\end{figure}

Seidr has four HMS-PLs, designed to correct wavefront errors associated with each of the VLTI beams. The HMS-PL design transforms the injected PSF at the MMF end of the lantern to six SMFs at the output. The central mode-selective core isolates the LP01 mode, for example, by having a larger diameter then the other cores, as demonstrated on the SMF output in Figure~\ref{fig:pl}. The wavefront sensing cores are then arranged in a pentagonal configuration around the central mode-selective core, also demonstrated at the SMF output in Figure~\ref{fig:pl}. The different colors represent different refractive indices, where darker blue represents a higher refractive index. The yellow arrows indicate the direction of light propagation.

The properties of the lantern simulated in this work are given in Table~\ref{tab:pl_param}. Of note, the mode-selective core has a radius double that of the wavefront sensing cores (representing SMF-28~\cite{Kowalevicz2006}), while the refractive indices remain the same. 

\begin{table}[ht]
    \caption{Photonic lantern properties.}
    \label{tab:pl_param}
    \begin{center}
    \begin{tabular}{|l|c|}
        \hline
        \rule[-1ex]{0pt}{3.5ex} \textbf{Parameter} & \textbf{Value} \\ \hline
        \rule[-1ex]{0pt}{3.5ex} PL length & 50,000~$\mu$m \\ \hline
        \rule[-1ex]{0pt}{3.5ex} WFS SMF core radii & 4.1~$\mu$m \\ \hline
        \rule[-1ex]{0pt}{3.5ex} MS SMF core radius & 8.2~$\mu$m \\ \hline
        \rule[-1ex]{0pt}{3.5ex} Output cladding radius & 155~$\mu$m \\ \hline
        \rule[-1ex]{0pt}{3.5ex} Output core spacing & 102.5~$\mu$m \\ \hline
        \rule[-1ex]{0pt}{3.5ex} Taper ratio & 20 \\ \hline
        \rule[-1ex]{0pt}{3.5ex} WFS core refractive indices & 1.449 \\ \hline
        \rule[-1ex]{0pt}{3.5ex} MS core refractive indices & 1.449 \\ \hline
        \rule[-1ex]{0pt}{3.5ex} Cladding refractive index & 1.444 \\ \hline
        \rule[-1ex]{0pt}{3.5ex} Capillary refractive index & 1.435 \\ \hline
        \end{tabular}
    \end{center}
\end{table}

 Light propagation through the lantern is a linear transformation of the complex electric field from the MMF input to the SMF output, where the entire transformation can be represented by the single complex-valued transfer matrix $\mathcal{A}$, such that the input electric field $\mathbf{u}_{in}$ can be transformed into the output electric field $\mathbf{u}_{out}$ as, ${\mathbf{u}_{out} = \mathcal{A} \mathbf{u}_{in}}$. Given the input vector $\mathbf{u}_{in}$ of length $m$ at the MMF end of the lantern and output vector $\mathbf{u}_{out}$ of length $n$ at the SMF end of the lantern, the transfer matrix $\mathcal{A}$ has dimensions $n \times m$. 
 
\begin{figure}[ht] 
    \begin{center}
    \includegraphics[scale=0.6]{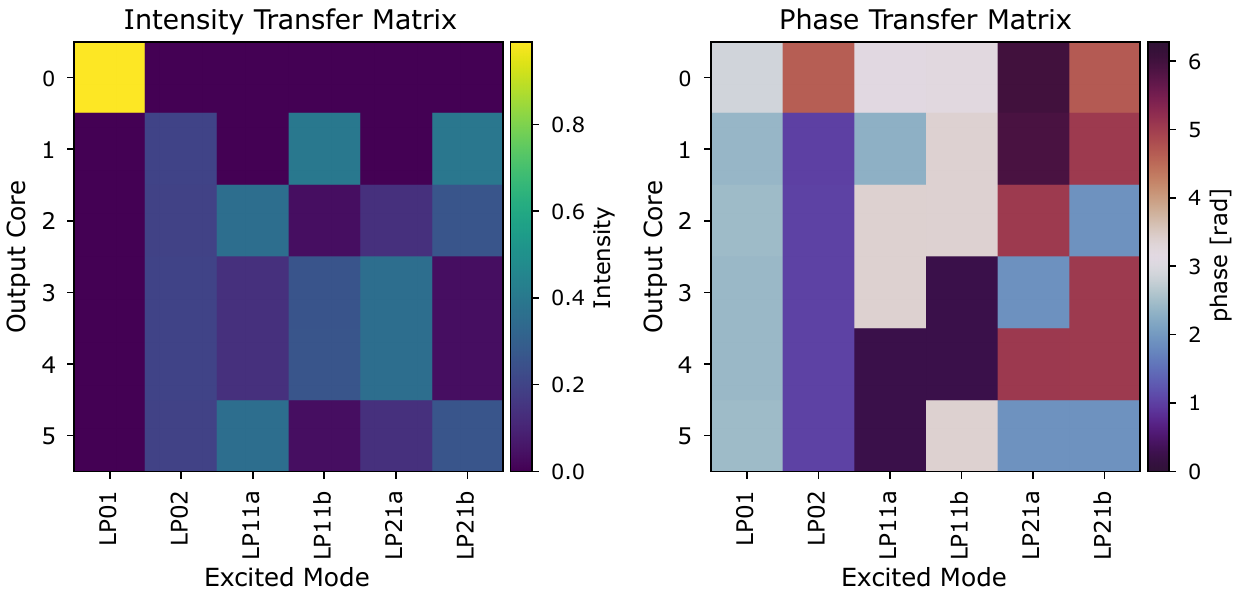}
    \end{center}
    \caption{6-core hybrid mode-selective photonic lantern intensity and phase transfer functions for $\lambda=1550$~nm.}
    \label{fig:tf}
\end{figure}

 Figure~\ref{fig:tf} describes the transfer matrix implemented in this work, where the lantern defined in Table~\ref{tab:pl_param} is characterized using a finite-difference beam propagation method~\cite{Pedrola2015} with adaptive meshing~\cite{Shibayama1999}. It can be seen that the LP01 mode is captured by the central core~0, while the higher-order modes are captured by the wavefront sensing cores~1~to~5, as desired. Note that only six modes are guided by the MMF end.

Only the intensities of the SMFs are measured at the output, such that the vector of output intensities, ${\mathbf{P}_{out} = [P_0, P_1, P_2, P_3, P_4, P_5]}$, is calculated as, ${\mathbf{P}_{out} = |\mathcal{A} \mathbf{u}_{in}|^2}$. For wavefront sensing, the objective is to map $\mathbf{P}_{out}$ to a wavefront estimation to compensate for the pupil-plane wavefront distortions.

\begin{figure}[ht] 
    \begin{center}
    \includegraphics[scale=0.62]{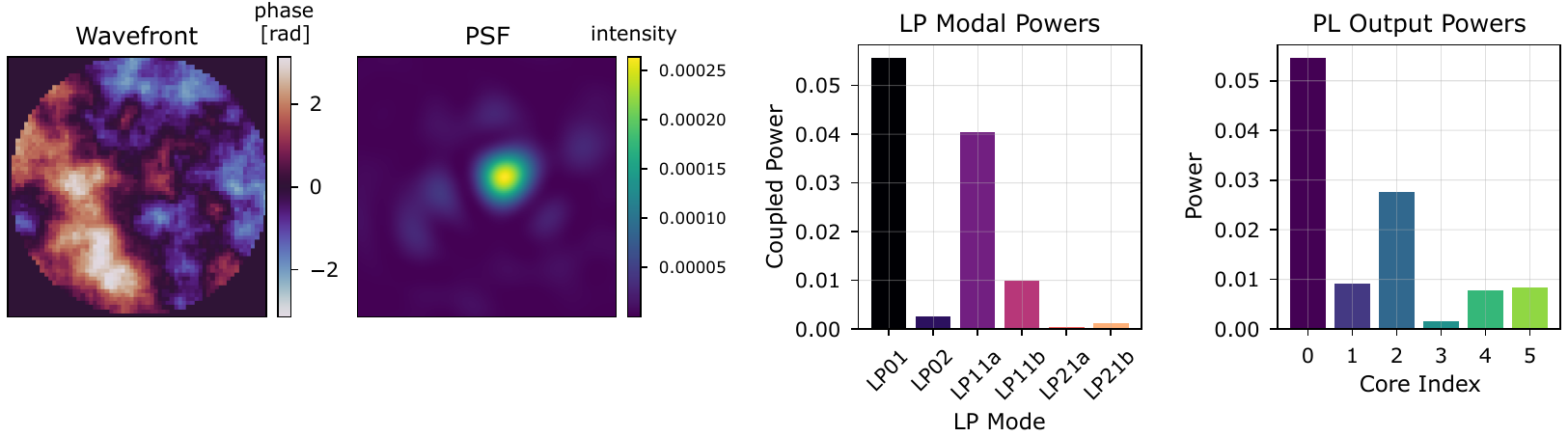}
    \end{center}
    \caption{Example of the simulations from a phase screen, to the focal plane PSF, to the injected LP modes, and the output lantern intensities.}
    \label{fig:seidr_sim}
\end{figure}

Figure~\ref{fig:seidr_sim} gives a single example of a simulated phase screen, PSF, decomposed LP intensities, and output lantern intensities.

The objective of the HMS-PLs in Seidr is to sense the incoming wavefront, then estimate a wavefront correction which would maximize the light in the LP01 mode, thus minimizing the light in the rest of the LP modes, resulting in maximal light in core~0 at the output of the lantern.

\section{WAVEFRONT ESTIMATION ALGORITHM} \label{sec:wfc}

Seidr\textquoteright s HMS-PLs act as both wavefront sensors and as optimal light injectors into the kernel-nulling chip. A wavefront estimation algorithm is designed to take the lantern SMF core intensities as input, then output an estimate of the wavefront correction to apply to Baldr\textquoteright s upstream DM with some static beam dependent offloading of tip-tilt to fold mirrors in the BIFROST injection optics.

In this work, we briefly describe the current iteration of the open-loop wavefront estimation algorithm designed for Seidr, please refer to~\cite{Long2026_seidrml} for an in-depth description of wavefront estimation for HMS-PLs.

A transformer NN (TNN) architecture is designed as the wavefront estimation algorithm for Seidr. Essentially, the TNN can take a sequential time-series as input, then estimate a wavefront correction to apply for the current time step (in the future, a wavefront prediction for the next time step). The TNN is trained using known sequential lantern intensities as inputs, and known wavefronts as outputs. During operation, sequential measured lantern intensities are input to the TNN, which outputs an estimate of the wavefront correction for the current time step. 

\begin{figure}[ht] 
    \begin{center}
    \includegraphics[scale=0.6]{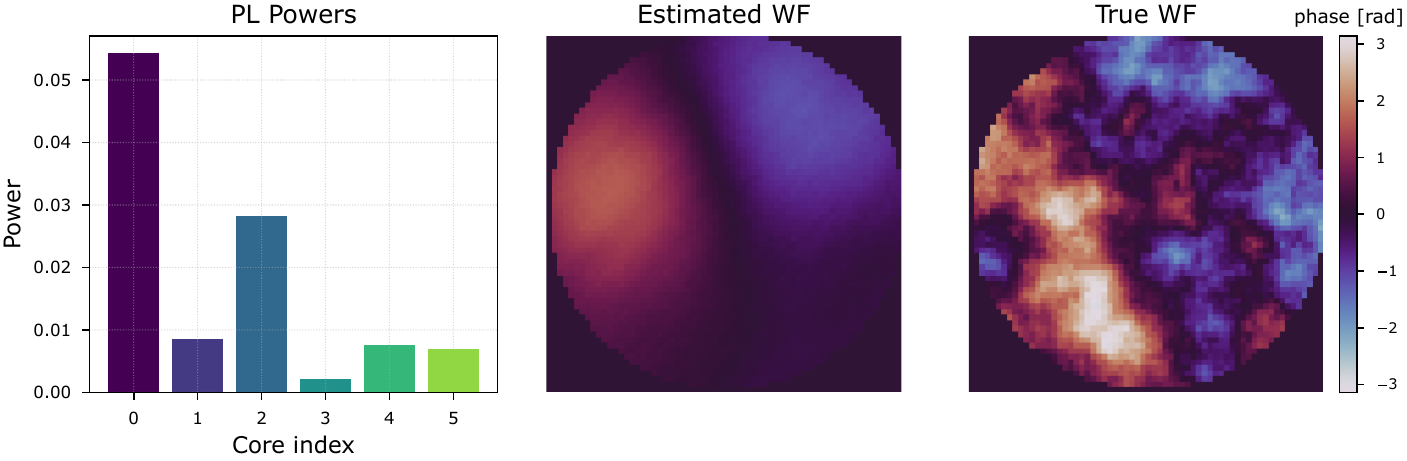}
    \end{center}
    \caption{Example of the wavefront estimation algorithm, with photonic lantern (PL) intensities input to a transformer neural network (TNN), an estimate of the wavefront output from the TNN, and the true wavefront being estimated. The recovered wavefront is consistent with the true values in the low order modes to which this photonic lantern is sensitive.}
    \label{fig:tnn_est}
\end{figure}

Using Von K\'arm\'an seeing as a test case~\cite{Schmidt2010}, Figure~\ref{fig:tnn_est} shows an example of lantern intensities used as the input to a trained TNN, the TNN output wavefront estimation, and the true wavefront being estimated.

\begin{figure}[ht] 
    \begin{center}
    \includegraphics[scale=0.5]{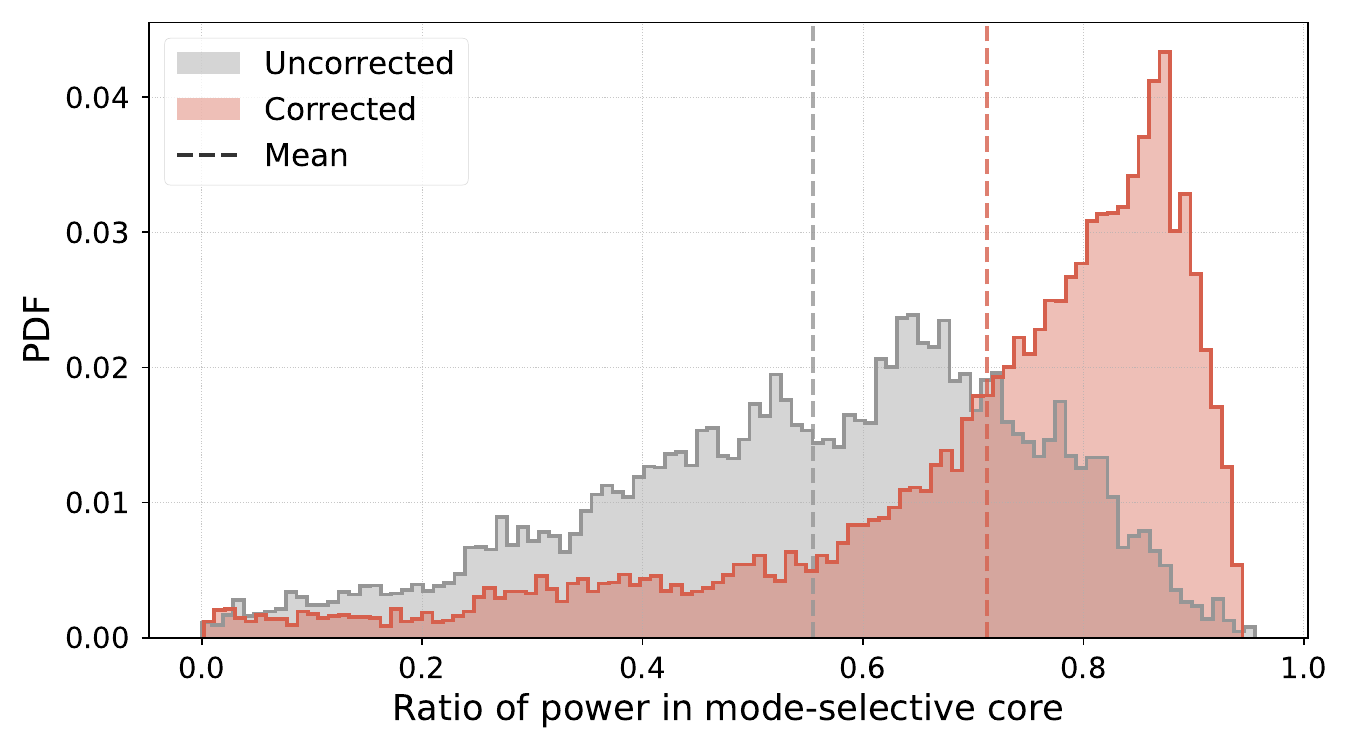}
    \end{center}
    \caption{Probability density function (PDF) of the ratio of power in the mode-selective core, before and after open-loop wavefront correction using the transformer neural network. The ratio of power in the mode-selective core increases after the TNN estimated wavefront correction is applied, with the average increasing from~0.55~to~0.71 (dashed lines).}
    \label{fig:mode_light}
\end{figure}

Focusing on the objective to maximize light in the mode-selective core for injection into the kernel-nulling chip, we provide an example of the probability density function (PDF) of the ratio of power in the mode-selective core for the uncorrected wavefront and the corrected wavefront in Figure~\ref{fig:mode_light}.

The results in Figure~\ref{fig:mode_light} are shown for Von K\'arm\'an seeing, with a Fried parameter of 0.4~m and an outer scale of 10~m, where Taylor\textquoteright s frozen turbulence hypothesis is used to generate temporal sequences with a transverse wind speed of 10~m/s~\cite{Taylor1938}. 

The uncorrected wavefront resulted in a mean ratio of 0.55, while the corrected wavefront resulted in a mean ratio of 0.71, a 28$\%$ increase in light injected into the mode-selective core. We note that the ratio of power in the mode-selective core does not take into account the overall amount of light injected into the kernel-nulling chip, and does not represent a closed-loop AO system. Instead, we aim to demonstrate the advantage of coupling the HMS-PLs upstream of the kernel-nulling chip to enhance instrument performance.

In operation, there exists a unique challenge in wavefront estimation for the purpose of maximizing light injection into the mode-selective core of a HMS-PL --- increasing the ratio of light injected into the mode-selective core decreases the amount of light in the rest of the wavefront sensing cores; thereby decreasing the signal-to-noise ratio, which subsequently affects the wavefront sensing performance. We address this problem and potential solutions further in~\cite{Long2026_seidrml}.

\section{KERNEL-NULLING} \label{sec:knc}

After passing through the upstream VLTI and Asgard optics, light from the HMS-PL mode-selective cores is injected into the kernel-nulling chip. Kernel-nulling is a two-stage nulling architecture, which is robust to phase aberrations~\cite{Chingaipe2023_kernel}. For Seidr, we propose an architecture with three observable quantities, known as \textit{kernels}. 

\begin{figure*}[ht]
    \centering
    \includegraphics[scale=1.0]{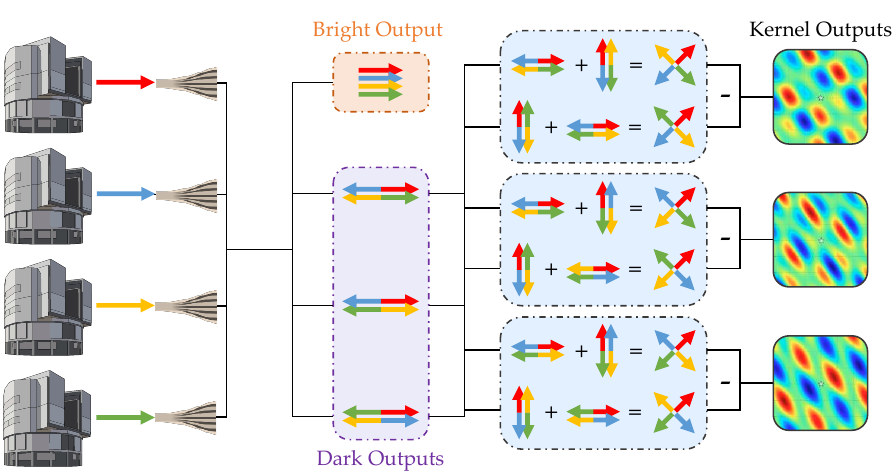}
    \caption{Kernel-nulling chip schematic, with four input beams passing through the VLTI and Asgard to Seidr. The nulling stage has one bright output and three dark outputs (six combinations), which are used to construct three kernel observables.}
    \label{fig:knc}
\end{figure*}

Figure~\ref{fig:knc} provides a schematic of a kernel-nulling chip architecture with three observable kernels. After passing through the HMS-PLs, the input electric field of each beam first propagates through a nulling stage, followed by a sensing stage.

The nulling stage produces three dark outputs and one bright output, then a subsequent sensing stage forms two non-symmetric combinations from each nuller output, resulting in six combinations of dark outputs. Without considering the temporal phase fluctuations, and ignoring the bright output from the nuller, the nulling and sensing operators can be combined into the mixing operator $\mathbf{M}$ as,
\begin{equation} \label{eq:mat_M}
    \mathbf{M} = \frac{1}{{4}} \times 
    \begin{bmatrix} 1 + e^{i\theta} & 1 - e^{i\theta} & -1 + e^{i\theta} & -1 - e^{i\theta} \\ 
    1 - e^{-i\theta} & -1 - e^{-i\theta} & 1 + e^{-i\theta} & -1 + e^{-i\theta} \\ 
    1 + e^{i\theta} & 1 - e^{i\theta} & -1 - e^{i\theta} & -1 + e^{i\theta} \\ 
    1 - e^{-i\theta} & -1 - e^{-i\theta} & -1 + e^{-i\theta} & 1 + e^{-i\theta} \\
    1 + e^{i\theta} & -1 - e^{i\theta} & 1 - e^{i\theta} & -1 + e^{i\theta} \\
    1 - e^{-i\theta} & -1 + e^{-i\theta} & -1 - e^{-i\theta} & 1 + e^{-i\theta} \end{bmatrix},
\end{equation}

\noindent where $\theta$ is a pre-defined phase offset. The mixing operator takes in the light from four telescopes, then outputs six nulled complex amplitudes. A detector then measures the six intensities, represented by the vector ${\mathbf{x} = |\mathbf{M} \cdot \mathbf{E}|^2}$. For a nuller with six outputs, three kernels can be assembled (equating the independent closure phases for a four input interferometer). When all of the incoming beams are in-phase, the system is nulled, and the first-order derivatives of the phase and amplitude are all zeros. 

Further, the second-order instrumental phase errors are eliminated via the three linear nuller output combinations. Defining the second-order phase errors as the matrix $\mathbf{\Theta}$, a kernel operator $\mathbf{K}$ can be defined, such that the second-order instrumental phase errors are eliminated via ${\mathbf{K} \cdot \mathbf{\Theta} = 0}$. The kernel observables $\mathbf{y}$ are then calculated as ${\mathbf{y} = \mathbf{K} \cdot \mathbf{x}}$. The independence of the kernel outputs to second-order phase perturbations is well suited to high contrast imaging~\cite{Martinache2018_kernel, Chingaipe2023_kernel}.

In practice, there exists temporal relative phase and intensity noise fluctuations, due to distortion of the light passing through the atmosphere, upstream optics, and detector noise. Intensity stabilization is provided by the HMS-PLs, while Heimdallr provides phase stabilization using its fringe-tracker. However, deeper nulls could be achieved using a series of on-chip phase shifters, which are planned for a second generation of the Seidr instrument.

\begin{figure*}[ht]
    \centering
    \includegraphics[scale=0.65]{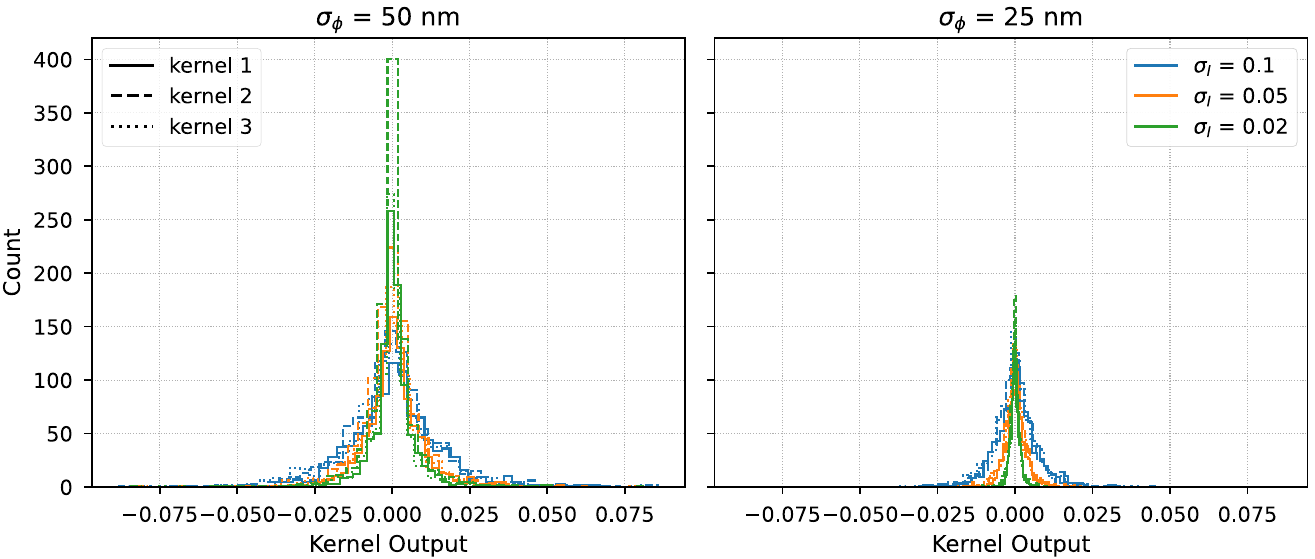}
    \caption{Example of kernel outputs for residual phase errors of $\sigma_{\phi}=50$~nm and $\sigma_{\phi}=25$~nm, given different relative intensity noise fluctuations: ${\sigma_I=0.10}$, ${\sigma_I=0.05}$, and ${\sigma_I=0.02}$. ${\sigma_I=0.10}$ represents the kernel outputs without the hybrid mode-selective photonic lanterns (HMS-PLs), while ${\sigma_I=0.05}$ and ${\sigma_I=0.02}$ show potential improvements in relative intensity noise fluctuations, as offered by the HMS-PLs.}
    \label{fig:knc_example}
\end{figure*}

To highlight the potential improvements in kernel-nulling performance offered by the HMS-PLs and on-chip phase shifters, we simulate light propagation through a kernel-nulling chip for different relative phase residual errors $\sigma_{\phi}$ and intensity noise fluctuations $\sigma_I$, showing their effects on the output kernels in Figure~\ref{fig:knc_example}. 

First, we set a fixed $\sigma_{\phi}$ from Heimdallr at $\sigma_{\phi}=50$~nm~\cite{Taras2024_seidr}, with no on-chip fringe-tracking. We note that this is a lower estimate of $\sigma_{\phi}$, given non-common-path vibrations and quasi-static wavefront aberrations between Heimdallr and Seidr, which are not sensed by Heimdallr\textquoteright s fringe-tracker. Second, we set $\sigma_{\phi}=25$~nm, assuming optimistic on-chip fringe-tracking could halve $\sigma_{\phi}$. We run 1000 simulations for both $\sigma_{\phi}$ values for three levels of intensity noise fluctuations $\sigma_I$: ${\sigma_I=0.10}$, ${\sigma_I=0.05}$, and ${\sigma_I=0.02}$. Here, we take ${\sigma_I=0.10}$ as representing post-AO (GPAO/NAOMI + Baldr) corrected beams in the H-band (e.g. under similar assumptions to~\cite{Martinache2018_kernel}). We then consider the effects of the HMS-PLs, which sense non-common-path wavefront errors, reducing the relative intensity fluctuations. We illustrate potential $\sigma_I$ improvements by factors of two (${\sigma_I=0.05}$) and five (${\sigma_I=0.02}$), respectively. 

For both $\sigma_{\phi}$ values, it can be seen that as $\sigma_I$ decreases, all three kernel output distributions narrow, while remaining centered on zero. Narrowing of the kernel outputs is more distinct as $\sigma_{\phi}$ reduces, which is due to both the residual phase-only uncertainty of the kernel outputs and the cross-term uncertainty between the residual phase errors and intensity fluctuations (see~\cite{Martinache2018_kernel} for a more in-depth discussion of the uncertainty terms affecting the kernel outputs). Narrowing of the kernel output distributions leads to a lowering of the noise floor of the null observable. As such, for a fixed integration time, reducing $\sigma_I$ results in improved detectable contrast limits, which becomes more pronounced as $\sigma_{\phi}$ is reduced. 

Overall, we show that Seidr is able to increase chip light injection and stabilization of the VLTI\textquoteright s beams by coupling its kernel-nulling chip with four HMS-PLs, offering higher contrast imaging of off-axis structures.

\section{CONCLUSION}
\label{sec:concl}

Seidr complements the Asgard Instrument Suite as an H-band kernel-nulling device. Seidr incorporates hybrid mode-selective photonic lantern modules for optimal light injection into its nulling chip, helping to improve contrast detection limits. Initial results show increased light injection into the nulling chip using our open-loop neural network-based wavefront estimation algorithm. Overall, this instrument update highlights the potential of the Seidr instrument for very high contrast and very high angular resolution imaging.



\acknowledgments 
 
We acknowledge support from Astralis - Australia\textquoteright s optical astronomy instrumentation Consortium - through the Australian Government\textquoteright s National Collaborative Research Infrastructure Strategy (NCRIS) Program. B.N. acknowledges support from an ARC Future Fellowship FTFT240100614. F.M acknowledges funding from the project PHOTONICS financed by the ANR program PEPR Origins (ANR-22-EXOR-0005).

\bibliography{bibliography} 
\bibliographystyle{spiebib} 

\end{document}